\documentclass[aps,twocolumn,pre,showpacs,preprintnumbers,eqsecnum,amssymb]{revtex4}

\usepackage{graphicx}
\usepackage{dcolumn}
\usepackage{bm}
\usepackage{amsmath}
\usepackage[T1]{fontenc}

\begin{document}

\preprint{arXiv-hyp-BIN-v3.tex}

\title{Hyperinflation in Brazil, Israel, and Nicaragua
revisited}

\author{Martin A. Szybisz}

\affiliation{Departamento de Econom\'{\i}a, Facultad
de Ciencias Econ\'omicas, Universidad de Buenos Aires,\\
Av. C\'ordoba 2122, RA--1120 Buenos Aires, Argentina}

\author{Leszek Szybisz}
 \altaffiliation{Corresponding author}
 \email{szybisz@tandar.cnea.gov.ar}
\affiliation{Laboratorio TANDAR, Departamento de F\'{\i}sica,
Comisi\'on Nacional de Energ\'{\i}a At\'omica,\\
Av. del Libertador 8250, RA--1429 Buenos Aires, Argentina}
\affiliation{Departamento de F\'{\i}sica, Facultad de
Ciencias Exactas y Naturales,\\
Universidad de Buenos Aires, Ciudad Universitaria,
RA--1428 Buenos Aires, Argentina}
\affiliation{Consejo Nacional de Investigaciones
Cient\'{\i}ficas y T\'ecnicas,\\
Av. Rivadavia 1917, RA--1033 Buenos Aires, Argentina}

\date{\today}

\begin{abstract}
The aim of this work is to address the description of
hyperinflation regimes in economy. The spirals of hyperinflation
developed in Brazil, Israel, and Nicaragua are revisited. This
new analysis of data indicates that the episodes occurred in
Brazil and Nicaragua can be understood within the frame of the
model available in the literature, which is based on a nonlinear
feedback (NLF) characterized by an exponent $\beta>0$. In the NLF
model the accumulated consumer price  index carries a finite time
singularity of the type $1/(t_c-t)^{(1-\beta)/\beta}$ determining
a critical time $t_c$ at which the economy would crash. It is
shown that in the case of Brazil the entire episode cannot be
described with a unique set of parameters because the time series
was strongly affected by a change of policy. This fact gives
support to the ``so called'' Lucas critique, who stated that
model's parameters usually change once policy changes.  On the
other hand, such a model is not able to provide any $t_c$ in the
case of the weaker hyperinflation occurred in Israel. It is shown
that in this case the fit of data yields $\beta \to 0$. This limit
leads to the linear feedback formulation which does not predict
any $t_c$. An extension for the NLF model is suggested.
\end{abstract}


\maketitle

There are $10^{11}$ stars in the galaxy. That used to be a huge number.
But it's only a hundred billion. It's less than the national deficit!
We used to call them astronomical numbers. Now we should call them
economical numbers.

{\bf Richard Feynman}

\section{Introduction}
\label{sec:introduction}

Inflation contribution is fundamental to reach the ``economical
numbers'' quoted by Feynman (see data plotted below). But most
importantly, when inflation surpasses moderate levels it affects real
economic activities. Models of hyperinflation are especially suitable
to emphasize that inflation implies bad ``states of nature'' in
economy. Wars, states bankruptcies, and changes of social regimens
are the characteristics of such regimens. These issues are analyzed
in textbooks on econophysics \cite{stanley99,stauffer99,sornette03b}.

The model for hyperinflation available in the literature is based on a
nonlinear feedback (NLF) characterized by an exponent $\beta>0$ of a
power law. In such an approach the accumulated consumer price index
(CPI) exhibits a finite time singularity of the form $1/(t_c-t)^{(1
-\beta)/\beta}$. This feature allows to determine a critical time $t_c$
at which the economy would crash. Although this model has been
successfully applied to many cases \cite{sornette03,szybisz08,%
szybisz09,szybisz10,szybisz15}, the authors of Ref.\ \cite{sornette03}
found difficulties in determining $t_c$ for regimes of hyperinflation
occurred in Brazil, Israel, and Nicaragua. Therefore, the present work
is devoted to revisit these episodes. It is shown that after a revision
of data is possible to predict reasonable values of $t_c$ for Brazil
and Nicaragua. However, in the case of Israel the difficulty persists,
this feature would be plausibly attributed to permanent but partial
efforts for stopping inflation. In order to follow better the evolution
of inflation we provide brief historical descriptions for these
countries.

The paper is organized in the following way. In Sec.\ \ref{sec:theory}
the NLF theory is outlined with details in order to present
self-consistently the tools applied for analyzing regimes of
hyperinflation. The episodes occurred in Brazil, Israel, and Nicaragua
are revisited in Sec.\ \ref{sec:revisiting}. Finally, Sec.\
\ref{sec:summary} is devoted to summarize conclusions.

\section{Theoretical background}
\label{sec:theory}

Let us recall that the rate of inflation $i(t)$ is defined as
\begin{equation}
i(t) = \frac{P(t)-P(t-\Delta t)}{P(t-\Delta t)}
= \frac{P(t)}{P(t-\Delta t)} - 1  \:, \label{infla}
\end{equation}
where $P(t)$ is the accumulated CPI at time $t$ and $\Delta t$ is the
period of the measurements. In the academic financial literature, the
simplest and most robust way to account for inflation is to take
logarithm. Hence, the continuous rate of change in prices is defined as
\begin{equation}
C(t) = \frac{\partial \ln{P(t)}}{\partial t} \:. \label{c_rate0}
\end{equation}
Usually the derivative of Eq.\ (\ref{c_rate0}) is expressed in a
discrete way as
\begin{eqnarray}
C(t+\frac{\Delta t}{2}) &=& \frac{\left[ \ln P(t+\Delta t)-
\ln P(t) \right]}{\Delta t} \nonumber\\
&=& \frac{1}{\Delta t}\,\ln \left[ \frac{P(t+\Delta t)}{P(t)} \right]
\:. \label{c_rate1}
\end{eqnarray}
The growth rate index (GRI) over one period is defined as
\begin{eqnarray}
r(t+\frac{\Delta t}{2}) &\equiv& C(t+\frac{\Delta t}{2})\,\Delta t
= \ln \left[ \frac{P(t+\Delta t)}{P(t)} \right] \nonumber\\
&=& \ln[1 + i(t+\Delta t)] = p(t+\Delta t) - p(t) \:, \nonumber\\
\label{rate1}
\end{eqnarray}
where a widely utilized notation
\begin{equation}
p(t) = \ln P(t) \:, \label{plog}
\end{equation}
was introduced. It is straightforward to show that the accumulated
CPI is given by
\begin{equation}
P(t) = P(t_0)\,\exp{\left[\frac{1}{\Delta t}\int^t_{t_0} r(t') dt'
\right]} \;. \label{pt}
\end{equation}

\subsection{Cagan's model of inflation}
\label{sec:cagan}

In his pioneering work, Cagan has proposed \cite{cagan56} a model of
inflation based on the mechanism of ``adaptive inflationary
expectation'' with positive feedback between realized growth of the
market price $P(t)$ and the growth of people's averaged expectation
price $P^*(t)$. These two prices are thought to evolve due to a
positive feedback mechanism: an upward change of market price $P(t)$
in a unit time $\Delta t$ induces a rise in the people's expectation
price $P^*(t)$, and such an anticipation pushes on the market price.
Cagan's assumptions may be cast into the following equations:
\begin{equation}
1 + i(t + \Delta t) = \frac{P(t+\Delta t)}{P(t)} = \frac{P^*(t)}{P(t)}
= \frac{P^*(t)}{P^*(t-\Delta t)} \:, \label{cag1}
\end{equation}
and
\begin{equation}
\frac{P^*(t+\Delta t)}{P^*(t)} = \frac{P(t)}{P(t-\Delta t)}
= 1 + i(t) \:. \label{cag2}
\end{equation}
Actually $P^*(t)/P(t)$ indicates that the process induces a non exact
proportional response of adaptation due to the fact that the expected
inflation $P^*(t)$ expands the response to the price level $P(t)$ in
order to forecast and meet the inflation of the next period. Now, one
may introduce the expected GRI
\begin{equation}
r^*(t+\frac{\Delta t}{2}) \equiv  C^*(t+\frac{\Delta t}{2})\,
\Delta t = \ln \left[ \frac{P^*(t+\Delta t)}
{P^*(t)} \right] \:. \label{rate2}
\end{equation}
So, expressions (\ref{cag1}) and (\ref{cag2}) are equivalent,
respectively, to
\begin{equation}
r(t+\frac{\Delta t}{2}) = r^*(t-\frac{\Delta t}{2}) \:, \label{rate3}
\end{equation}
and
\begin{equation}
r^*(t+\frac{\Delta t}{2}) = r(t-\frac{\Delta t}{2}) \:. \label{rate4}
\end{equation}
These relations imply
\begin{equation}
r(t+\Delta t)=r(t-\Delta t) \:, \label{cagan}
\end{equation}
giving a constant finite GRI equal to its initial value $r(t)=r(t_0)=
r_0$. The accumulated CPI evaluated using Eq.\ (\ref{pt}) leads to an
exponential law
\begin{equation}
P(t) = P_0\,\exp{\left[r_0\,\biggr(\frac{t-t_0}{\Delta t}\biggr)
\right]} \;, \label{pt0}
\end{equation}
where $P_0=P(t_0)$.

\subsection{Feedback contribution to the equation for inflation}
\label{sec:feedback}

Due to the fact that the CPI during spirals of hyperinflation grows
more rapidly than the exponential law given by Eq.\ (\ref{pt0}), the 
Cagan's model for inflation has been generalized by Mizuno, Takayasu, 
and Takayasu (MTT) \cite{mizuno02} including a linear feedback (LF)
process. For this purpose, the relation (\ref{cag1}) was kept, while
Eq.\ (\ref{cag2}) was replaced by
\begin{eqnarray}
\ln \left[ \frac{P^*(t+\Delta t)}{P^*(t)} \right]
&=& ( 1 + 2\,a_p)\,\ln \left[ \frac{P(t)}{P(t-\Delta t)} \right] \:,
\nonumber\\
\label{mizuno1}
\end{eqnarray}
which leads to
\begin{equation}
r(t+\Delta t) = r(t-\Delta t) + 2\,a_p\,r(t-\Delta t) \;.
\label{mizuno2}
\end{equation}
Here $a_p$ is a positive dimensionless feedback's strength, in fact,
MTT defined a parameter $B_{MTT}=1+2a_p$. In the continuous limit one
arrives at
\begin{equation}
\frac{dr}{dt} = \frac{a_p}{\Delta t}\,r(t)~~~~\to~~~~r(t) = r_0\,
\exp \biggr[a_p\,\biggr(\frac{t-t_0}{\Delta t}\biggr)\,\biggr] \;.
\label{r_mizu}
\end{equation}
In this approach the CPI grows as a function of $t$ following a
double-exponential law \cite{szybisz09,mizuno02}, so one gets
\begin{equation}
\ln P(t) = p(t) = p_0 + \frac{r_0}{a_p}\,\biggr\{ \exp \biggr[a_p\,
\biggr(\frac{t-t_0}{\Delta t}\biggr)\,\biggr] - 1 \biggr\} \;.
\label{p_mizu}
\end{equation}

Since in practice there are cases where $P(t)$ grows more rapidly than
a double-exponential law, in a next step, Sornette, Takayasu, and Zhou
(STZ) \cite{sornette03} included a nonlinear feedback process in the
formalism. In this approach, Eq.\ (\ref{cag1}) is also kept, whilst
Eq.\ (\ref{mizuno1}) is replaced by
\begin{eqnarray}
\ln \left[ \frac{P^*(t+\Delta t)}{P^*(t)} \right]
&=& \ln \left[ \frac{P(t)}{P(t-\Delta t)} \right] \nonumber\\
&\times& \left( 1 + 2\,a_p \biggr\{
\ln \left[\frac{P(t)}{P(t-\Delta t)}\right]\biggr\}^\beta \right) \:,
\nonumber\\
\label{rate00}
\end{eqnarray}
leading to
\begin{equation}
r(t+\Delta t) = r(t-\Delta t) + 2\,a_p\,[r(t-\Delta t)]^{1+\beta} \;.
\label{rate42}
\end{equation}
Here $\beta > 0$ is the exponent of the power law. In the discrete 
version of this NLF model, $r(t)$ follows a double-exponential law;
while $P(t)$ increases as a triple-exponential law \cite{szybisz08,%
szybisz09}. Notice that for $\beta=0$ this formulation retrieves the
LF proposal of MTT given by Eq.\ (\ref{mizuno2}).

Taking the continuous limit in Eq. (\ref{rate42}) one obtains the
following equation for the time evolution of $r$
\begin{equation}
\frac{dr}{dt} = \frac{a_p}{\Delta t}\,[r(t)]^{1+\beta} \;.
\label{rate43}
\end{equation}
For $\beta>0$ the solution for GRI follows a power law exhibiting a 
singularity at finite-time $t_c$ \cite{sornette03,szybisz08,szybisz09}
\begin{equation}
r(t) = r_0\,\biggr[\frac{1}{1-\beta\,a_p\,r_0^\beta
\left(\frac{t-t_0}{\Delta t}\right)} \biggr]^{1/\beta}
= r_0\,\left(\frac{t_c-t_0}{t_c-t}\right)^{1/\beta} \;. \label{r_time}
\end{equation}
The critical time $t_c$ being determined by the initial GRI $r(t=t_0)=
r_0$, the exponent $\beta$, and the strength parameter $a_p$
\begin{equation}
\frac{t_c - t_0}{\Delta t} = \frac{1}{\beta\,a_p\,r_0^\beta} \;.
\label{c_time}
\end{equation}
In turn, the $\log$-CPI for $\beta \ne 1$ is obtained by integrating
$r(t)$ according to Eq.\ (\ref{pt})
\begin{eqnarray} 
&&\ln \left[\frac{P(t)}{P_0}\right] = p(t) - p_0 = \int^t_{t_0} r(t')
\frac{dt'}{\Delta t} \nonumber\\
&=& \frac{r_0^{1-\beta}}{(1-\beta)\,a_p}
\biggr\{ \biggr[\frac{1}{1-\beta\,a_p\,r_0^\beta
\left(\frac{t-t_0}{\Delta t}\right)}
\biggr]^{\frac{1-\beta}{\beta}} - 1 \biggr\} \;. \nonumber\\
\label{lptn}
\end{eqnarray}
For $0<\beta<1$ one gets
\begin{equation}
p(t) = p_0 + \frac{r_0\,\beta}{1-\beta}
\left(\frac{t_c-t_0}{\Delta t}\right)
\biggr[\left(\frac{t_c-t_0}{t_c-t}\right)^{\frac{1-\beta}{\beta}}
- 1 \biggr] \;. \label{price1}
\end{equation}
This solution corresponds to a genuine divergence of $p(t)$, the
$\log$-CPI exhibits a finite-time singularity at the same critical
value $t_c$ as GRI. Let us emphasize that all the free parameters
have their own meaning: $t_c$ is the hyperinflation's end-point
time; $\beta$ is the exponent of the power law; $r_0$ is the
initial slope for the growth of $\log$-CPI; and $p_0$ is the
initial $\log$-CPI. Equation (\ref{price1}) has been used for the
analysis of hyperinflation episodes reported in previous papers
\cite{sornette03,szybisz09,szybisz08,szybisz10,szybisz15}.

\begin{figure*}
\includegraphics[width=8.5cm, height=6cm]{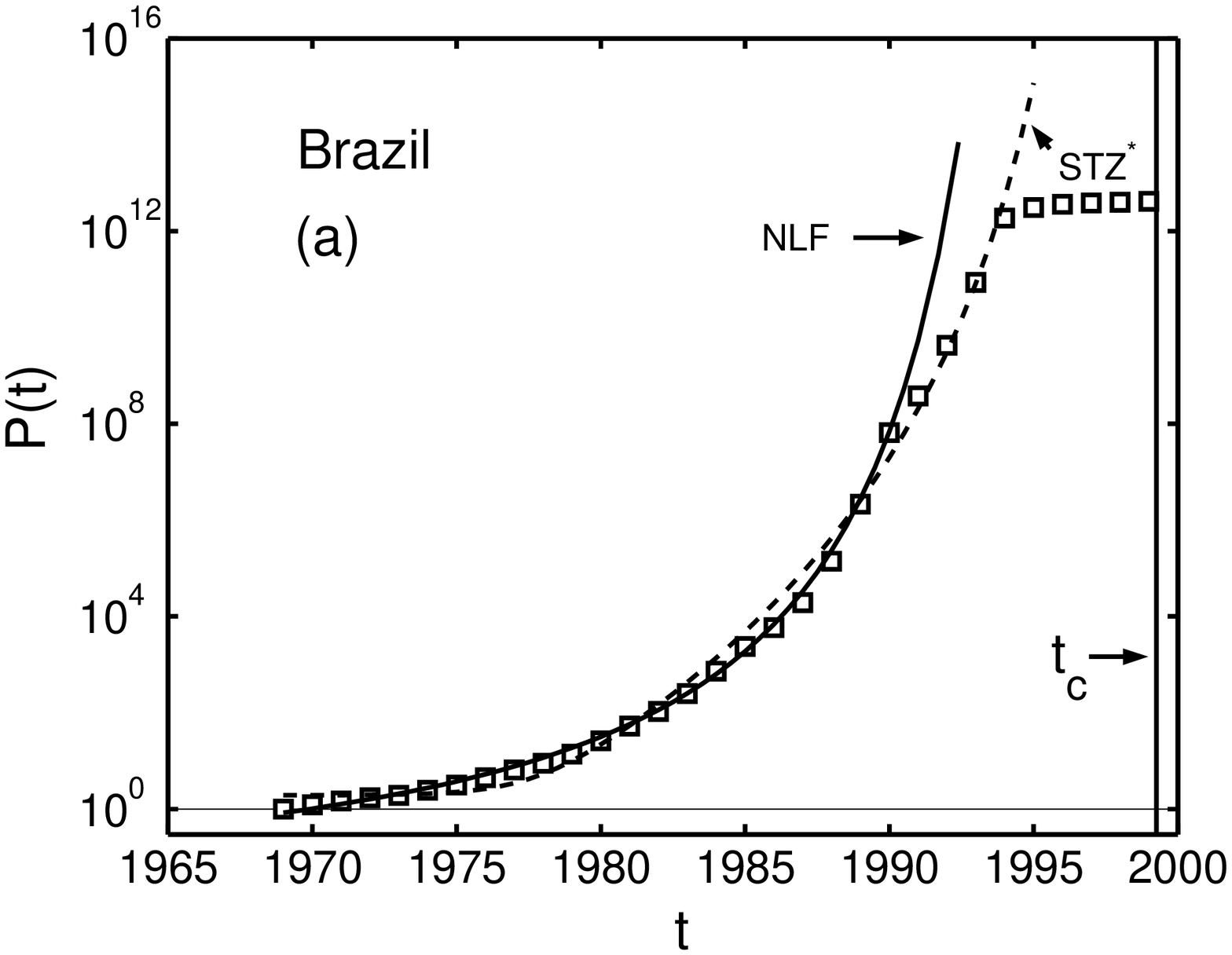}
\includegraphics[width=8.5cm, height=6cm]{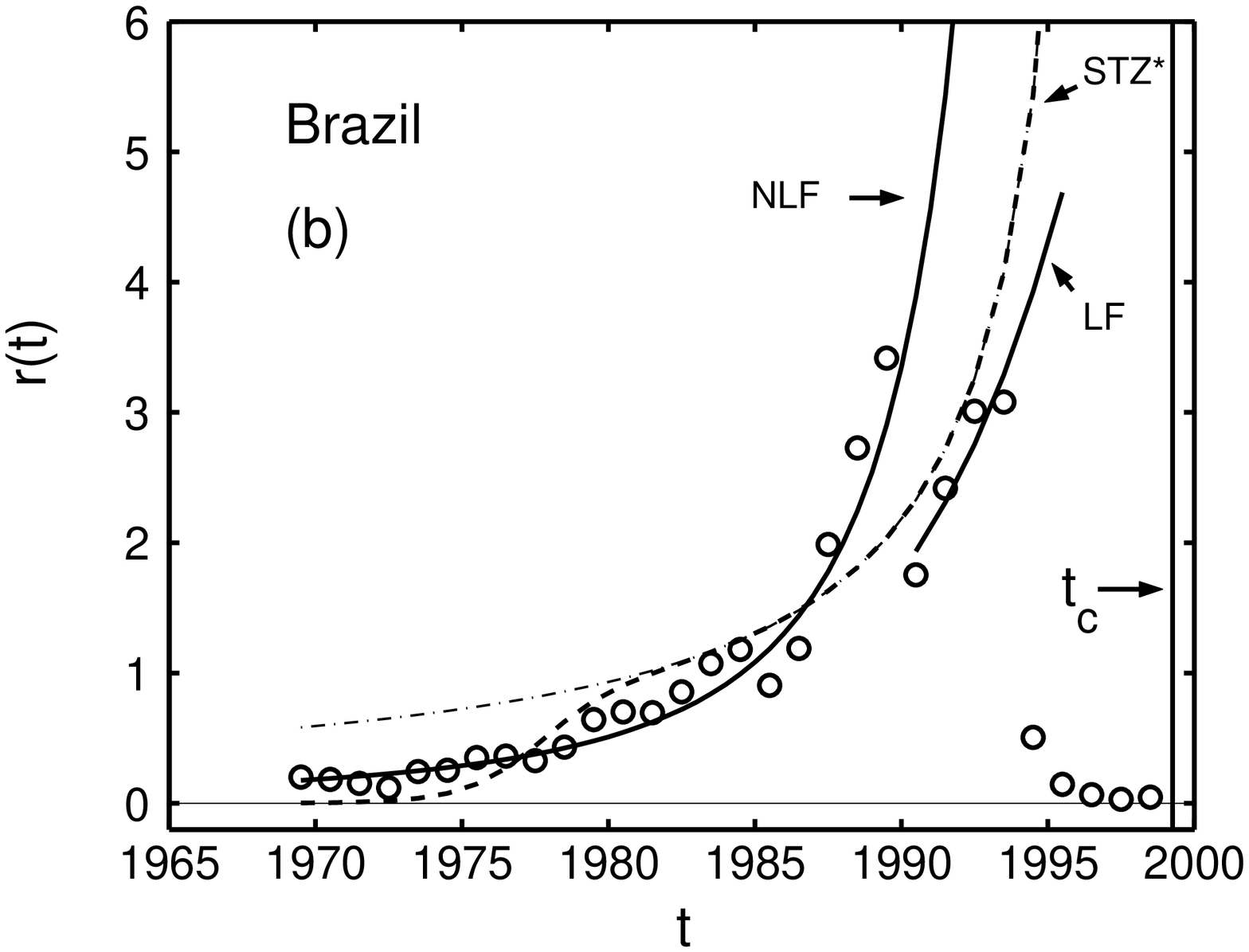}
\caption{\label{fig:Brazil_hyp}(a) Squares are yearly CPI in Brazil
from 1969 to 1999, normalized to $P(t_0=1969)=1$, presented in a
semi-logarithmic plot. (b) Circles are yearly GRI for the same period
as in (a). The solid curve in (a) is the fit of $\ln P(t)$ from 1969 to
1990 with Eq.\ (\ref{price1}), i.e. NLF model, while the dashed line is
the fit of $P(t)$ from 1969 to 1994 with Eq.\ (\ref{price1}) reported
by STZ* (see text). In panel (b) the solid curve NLF stands for $r(t)$
evaluated with Eq.\ (\ref{r_time}), while the solid curve LF is $r(t)$ 
evaluated with Eq.\ (\ref{r_mizu}), in both cases the parameters quoted
in Table \ref{tab:table1} were used. The dashed curve is $r(t)$
evaluated using Eq.\ (\ref{r_dP_1}) and the dot-dashed curve is the
asymptotic limit given by Eq.\ (\ref{r_dP_lim}) (see text). In both
panels the vertical solid line indicates the critical time $t_c$
predicted by data of the period 1969-1990.}
\end{figure*}

\section{Hyperinflation in Brazil, Israel and Nicaragua revisited}
\label{sec:revisiting}

We shall now revisit the episodes of hyperinflation developed in
Brazil, Israel, and Nicaragua performing a study within the framework
of the NLF model outlined in the previous section. These cases have
been already studied by Takayasu and collaborators \cite{mizuno02,%
sornette03}. In particular, in Ref.\ \cite{sornette03} the authors
stated: ``a fit of the price time series with expression (15) gives an 
exponent $\alpha$ larger than 15 and critical times $t_c$ in the range 
2020-2080, which are un-realistic'' ({\it sic}). Equation (15) of Ref.
\cite{sornette03} is equivalent to Eq.\ (\ref{price1}) of the present 
work and the parameter $\alpha$ written in terms of the exponent 
$\beta$ is
\begin{equation}
\alpha = (1-\beta)/\beta \;. \label{alfa}
\end{equation}
Hence, the results of Ref.\ \cite{sornette03} correspond to $\beta$
smaller than 0.07. In addition, the authors of Ref.\ \cite{sornette03}
said that the results are not improved by reducing the time intervals
over which the fits are performed. In seeking for how to overcome this 
problem, they found that reasonable critical $t_c$ are obtained after
a simple change of variable from $\ln P(t)$ to $P(t)$, i.e. by fitting
$P(t)$ instead of $\ln P(t)$ with Eq.\ (\ref{price1}).

\subsection{Hyperinflation of Brazil and Nicaragua}
\label{sec:failures}

We shall now proceed to discuss the entire regimes of hyperinflation
occurred in Brazil (1969-1994) and Nicaragua (1969-1991). In searching
why it was impossible to describe satisfactorily well these episodes
utilizing the NLF model the data of both these countries were revised.

Let us now present a short story of Brazilian economic difficulties.
In fact, Brazil was not defeated in a war nor was required to pay war
reparations, but the foreign debt accumulated in the 1970's by
borrowing large amounts of cheap petrodollars, the external shock of
1979 (second oil shock and interest rate shock) and the suspension of
new external financing since 1982 had together produced similar
consequences \cite{bresser91,merette00}. The country that in the 1970's
received around $2\%$ of gross domestic product (GDP) of foreign
savings was now required to transfer resources of 4 to $5\%$ to the
creditor countries. Debt service was equal to 83$\%$ of export earnings
in 1982. The country struggled to finance its external indebtedness and
growth came to a halt. These economic problems were accompanied by
political turbulence. The military dictatorship that had ruled Brazil
since 1964 lost support and was forced to step down in 1985, which
resulted in the return of democracy.

\begin{table*}
\caption{\label{tab:table1} Parameters obtained from the analysis of 
episodes of hyperinflation occurred in Brazil, Israel and Nicaragua.}
\begin{ruledtabular}
\begin{tabular}{llccccccll}
Country & Period & \multicolumn{6}{c}{Parameters} & Model & $\chi$ \\
\cline{3-8}
&& $t_c$ & $a_p$ & $r_0$ & $\beta$ & $\gamma$ & $p_0$ & & \\
\hline
 \\
Brazil    & 1969-1994 & 1997.50 & & 0.402$\times$10$^{-2}$ & 0.058
& & 1.93 & STZ$^a$ & 0.604 \\
          & 1969-1990 & 1999.26$\pm$6.22 & 0.172 & 0.165$\pm$0.029
& 0.383$\pm$0.152 &         &       0  & NLF & 0.190 \\
          & 1990-1994 & & 0.177$\pm$0.116 & 1.770$\pm$0.425
&   &   & 18.2 & LF & 0.158 \\

Israel    & 1969-1985 & 1988.06 & & 0.077 & 0.149  & & 1.04
& STZ$^a$ & 0.085 \\
          &           & & 0.176$\pm$0.035 & 0.101$\pm$0.035
&   & & 0 & LF & 0.088 \\
          &           & 2061$\pm$72 & 0.184 & 0.109$\pm$0.035
& 0.069$\pm$0.061 & & 0 & NLF & 0.095 \\
          &           & 2527$\pm$456 & 0.177 & 0.102$\pm$0.035
& 0.010$\pm$0.009 & & 0 & NLF & 0.089 \\
          & 1969-1984 & & 0.178$\pm$0.045 & 0.100$\pm$0.040
&   & & 0 & LF & 0.089 \\
          &           & 2048$\pm$79 & 0.189 & 0.107$\pm$0.041
& 0.080$\pm$0.093 & & 0 & NLF & 0.094 \\
          &           & 2588$\pm$619 & 0.179 & 0.101$\pm$0.040
& 0.009$\pm$0.010 & & 0 & NLF & 0.090 \\

Nicaragua & 1969-1991$^b$ & 1992.91 && 0.881$\times$10$^{-5}$
& 0.063 & & 3.24 & STZ$^a$ & 0.848 \\
          & 1969-1987$^c$ & 1987.71$\pm$0.87 & 0.383
& 0.101$\pm$0.031 & 0.710$\pm$0.217 &          & 0 & NLF & 0.298 \\
          & 1969-1988$^c$ & 1992.32$\pm$2.38 & 0.316
& 0.067$\pm$0.020 & 0.356$\pm$0.102 &          & 0 & NLF & 0.519 \\

\end{tabular}
$^a$ The values of the parameters listed in this line were calculated
using those reported by STZ \cite{sornette03} obtained by fitting
data of $P(t)$ instead of $\ln P(t)$ to Eq.\ (\ref{price1}), see
text. \\
$^b$ Data from Ref.\ \cite{imf}. \\
$^c$ Data from Ref.\ \cite{imf_2011}.
\end{ruledtabular}
\end{table*}

Since its inauguration in January 1985, the first democratic government
after military rule exerted by the elected vice-president Jos\'e Sarney
(because the elected president Tancredo Neves fell ill) had
limited means to resist spending pressure from congress. As a result,
inflation, which had already been high for several years thanks to the
old practice of monetary financing of budget deficits, frequent
devaluations and indexation (automatic correction of prices, interest
rates and wages according to past inflation), ran totally out of
control. In 1987, the government was not able to pay the interest on 
its foreign debt and Brazil's public debt had to be rescheduled. The 
inflation peaked at 2,950 $\%$ in 1990. This behavior can be seen in 
Fig.\ \ref{fig:Brazil_hyp}(b), where data of yearly GRI computed using
values taken from a Table of the International Monetary Fund (IMF)
\cite{imf} are displayed. A new elected president Fernando Collor de
Mello applied in 1990 the so-called Collor's Plan in order to stop
hyperinflation. As can be seen in Fig.\ \ref{fig:Brazil_hyp}(b) at
the beginning the trend was changed, however, finally this plan for
stabilization failed \cite{bresser91,merette00}.

The launch of the Plano Real in 1994 would prove to be the turning 
point. This plan, designed by Henrique Cardoso, who would later become 
Brazil's president, envisaged the introduction of a new currency, put 
constraints on public spending and ended the indexation of the economy.
The new currency, the real, had a crawling peg against the dollar as a 
nominal anchor and was somewhat overvalued, which made imports cheap, 
thus limiting the room for domestic producers to raise prices.

Sornette, Takayasu, and Zhou \cite{sornette03} have analyzed the 
complete series of CPI data from 1969 to 1994. The results from the
fit of $P(t)$, instead of values of  $\ln P(t)$, with the 
right-hand-side (r.h.s.) of expression (\ref{price1}) reported by STZ 
in Table 2 of Ref.\ \cite{sornette03} are quoted in Table
\ref{tab:table1} and displayed in Fig.\ \ref{fig:Brazil_hyp}(a). For 
the sake of completeness, we provide the relations between the 
parameters $\alpha$, $A$, and $B$ utilized by STZ and that used in the 
present work
\begin{eqnarray}
&& r_0/\Delta t = \alpha\,B/(t_c-t_0)^{1+\alpha} \;, \label{r0}
 \\
&& \beta = 1/(1+\alpha) \;, \label{bet}
 \\
&& p_0 = A + B/(t_c-t_0)^\alpha \;. \label{p0}
\end{eqnarray}

We believe that the difficulty for fitting $\ln P(t)$ with Eq.\
(\ref{price1}) arises from the fact that in 1991 there is an important
departure from the initial trend clearly depicted in Fig.\
\ref{fig:Brazil_hyp}(b). The applied theory with a unique value of
$\beta$ is not able to describe the entire process. This feature is in
agreement with the statement of Lucas \cite{lucas76} that parameters
can change when economic policy changes. Therefore, in the present work
we fitted to Eq.\ (\ref{price1}) the data of $\ln P(t)$ previous to
1991 only. Preliminary results have been already reported in Ref.\
\cite{szybisz10}. The numerical task was accomplished by using a
routine of the book by Bevington \cite{bevington} cited as the first
reference in Chaps. 15.4 and 15.5 of the more recent {\it Numerical
Recipes} \cite{press96}. In practice, the applied procedure yields the
uncertainty in each parameter directly from the minimization algorithm.
In order to quantify the contribution of the feedback we also evaluated
$a_p$, which is given by [see Eq.\ (\ref{c_time})]
\begin{equation}
a_p = \frac{\Delta t}{\beta\,r_0^\beta\,(t_c-t_0)} \;. \label{a_p_lim}
\end{equation}
The obtained parameters, its uncertainties and the root-mean-square
(r.m.s) residue of the fit, i.e. $\chi$, are quoted in Table
\ref{tab:table1}. The determined $t_c$ is quite reasonable and the good
quality of this fit may be observed in Fig.\ \ref{fig:Brazil_hyp}(a).
The GRI was calculated by using Eq.\ (\ref{r_time}) and displayed in
Fig.\ \ref{fig:Brazil_hyp}(b), the theoretical results follows quite
good the measured data. Vertical lines in both (a) and (b) panels
indicate the obtained critical time $t_c$.

\begin{figure}
\includegraphics[width=8.5cm, height=6cm]{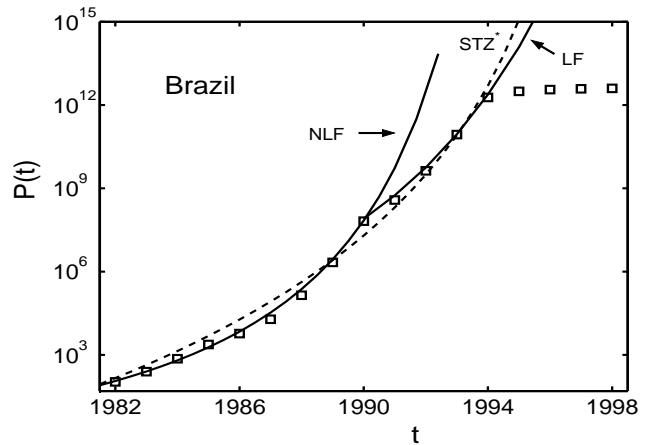}
\caption{\label{fig:Brazil_fork} Squares are yearly CPI in Brazil from
1982 to 1998, normalized to $P(t_0=1969)=1$, presented in a
semi-logarithmic. The solid curve NLF is the fit of $\ln P(t)$ from
1969 to 1990 with Eq.\ (\ref{price1}), while the solid curve LF stands
for the fit of $\ln P(t)$ from 1990 to 1994 with Eq.\ (\ref{p_mizu}).
The dashed line is the fit of $P(t)$ reported by STZ* (see text).}
\end{figure}

Figure \ref{fig:Brazil_fork} clearly shows a bifurcation between the
trend of data from 1969 to 1990 and that of data from 1990 to 1994. 
Since there are only a few data points for the new incipient branch of 
hyperinflation, in order to have a quantitative description data of 
$r(t)$ and $\ln P(t)$ from 1990 to 1994 were simultaneously fitted 
with Eqs.\ (\ref{r_mizu}) and (\ref{p_mizu}), respectively. The 
obtained parameters are included in Table \ref{tab:table1} and the 
fits denoted by LF are displayed in Figs.\ \ref{fig:Brazil_hyp}(b) and 
\ref{fig:Brazil_fork}.

On the other hand, one may observe in Fig.\ \ref{fig:Brazil_fork} 
that the fit reported by STZ* does not follow quite well the set of 
measured CPI. The situation is even worse when one examine the GRI.
According to the statement quoted on the top of page 499 of Ref.\
\cite{sornette03} the accumulated $P(t)$ is the exponential of the
integral of $r(t)$ as expressed in Eq.\ (\ref{pt}) of the present work.
The inverse, i.e. $r(t)$, becomes
\begin{equation}
r(t) = \frac{d \ln P(t)}{d(t/\Delta t)} = \frac{1}{P(t)}\,
\frac{d P(t)}{d(t/\Delta t)} \;. \label{r_dP_0}
\end{equation}
Assuming that $P(t)$ is given by Eq.\ (\ref{price1}) one gets
\begin{equation}
r(t) = \frac{r_0\,\left(\frac{t_c-t_0}{t_c-t}\right)^{1/\beta}}{p_0 +
\frac{r_0\,\beta}{1-\beta}\,\left(\frac{t_c-t_0}{\Delta t}\right)
\biggr[\left(\frac{t_c-t_0}{t_c-t}\right)^{\frac{1-\beta}{\beta}}
- 1 \biggr]} \;. \label{r_dP_1}
\end{equation}
An evaluation of $r(t)$ by using this formula with the corresponding
parameters quoted in Table \ref{tab:table1} yielded the dashed curve
depicted in Fig.\ \ref{fig:Brazil_hyp}(b). The theoretical curve
oscillates between both branches of measured data. It is interesting
to notice that the asymptotic form of Eq.\ (\ref{r_dP_1}) for $t \to
t_c$ becomes
\begin{equation}
r_{asympt}(t) = \frac{1-\beta}{\beta}\,\left(\frac{\Delta t}{t_c-
t}\right) \;, \label{r_dP_lim}
\end{equation}
yielding a universal singularity $(t_c -t)^{-1}$. This asymptotic
regime is reached quite soon as shown by the dot-dashed curve in Fig.\
\ref{fig:Brazil_hyp}(b).

\begin{table}
\caption{\label{tab:Nicaragua} Inflation in Nicaragua 1980-1997.}
\begin{ruledtabular}
\begin{tabular}{lrr}
Year &  \multicolumn{2}{c}{Annual averaged $i(\%)$} \\
\cline{2-3}
     & IMF-1$^a$ & IMF-2$^b$ \\
\hline
1980 &    35.1 &    35.1  \\
1981 &    23.8 &    23.8  \\
1982 &    24.9 &    28.5  \\
1983 &    31.1 &    33.6  \\
1984 &    35.4 &   141.3  \\
1985 &   219.5 &   571.4  \\
1986 &   681.0 &   885.2  \\
1987 &   911.9 & 13109.5  \\
1988 & 14315.8 &  4775.2  \\
1989 &  4709.3 &  7428.7  \\
1990 &  3127.5 &  3004.1  \\
1991 &  7755.3 &   116.6  \\
1992 &    40.5 &    21.9  \\
1993 &    20.4 &    13.5  \\
1994 &     7.7 &     3.7  \\
1995 &    11.2 &    11.2  \\
1996 &    11.6 &    11.6  \\
1997 &     9.2 &     9.2  \\
\end{tabular}
$^a$ Data from IMF World Economic Outlook (WEO) \cite{imf}. \\
$^b$ Data from IMF (WEO) \cite{imf_2011}.
\end{ruledtabular}
\end{table}

Let us now refer to the hyperinflation occurred in Nicaragua. During
the decade of 1970's this country was involved in a civil war. In 1979,
the Sandinista National Liberation Front (FSLN) overthrew Anastasio
Somoza Debayle, ending the Somoza dynasty, and established a
revolutionary government in Nicaragua. This new government, formed in
1979 and dominated by the FSLN, applied a new model of economic
development. The leader of this administration was Daniel Jos\'e Ortega
Saavedra (Sandinista junta coordinator 1979-85, president 1985-90).

\begin{figure*}
\includegraphics[width=8.5cm, height=6cm]{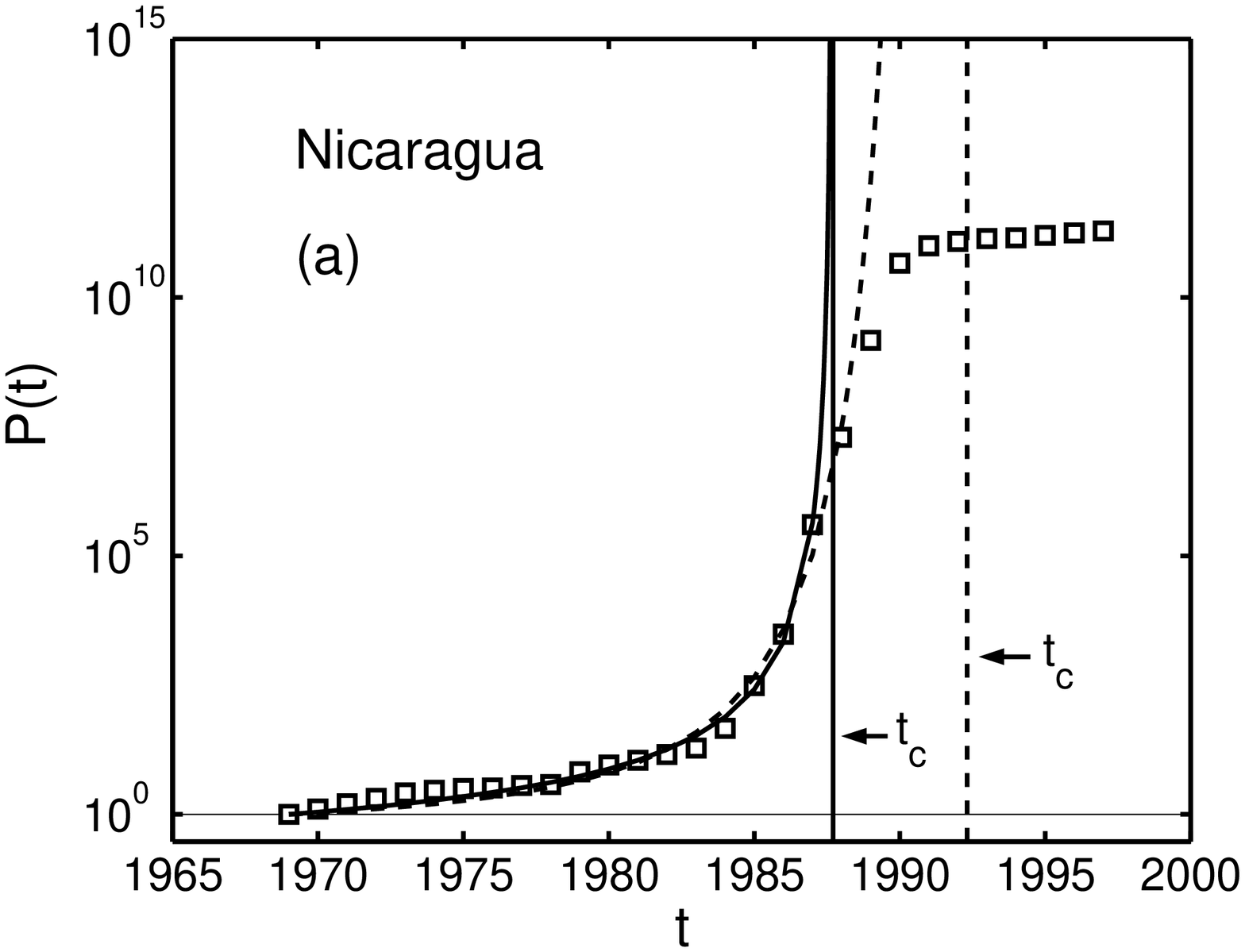}
\includegraphics[width=8.5cm, height=6cm]{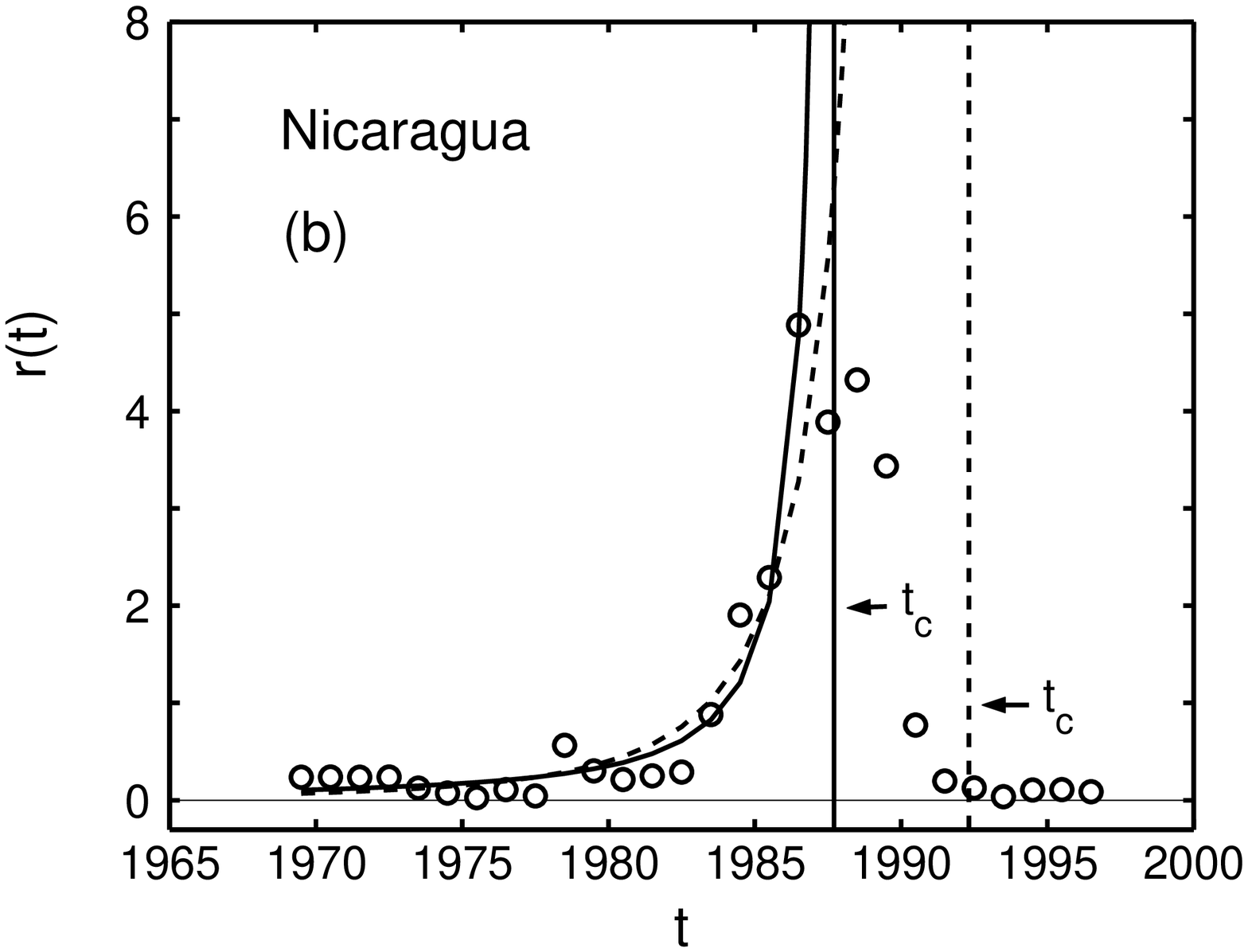}
\caption{\label{fig:Nicaragua}(a) Squares are yearly CPI in Nicaragua
from 1969 to 1997, normalized to $P(t_0=1969)=1$, presented in a
semi-logarithmic plot. (b) Circles are yearly GRI in Nicaragua for the
same period as in (a). The solid curve in (a) is the fit of $\ln P(t)$
from 1969 to 1987 with Eq.\ (\ref{price1}), NLF model, while the dashed
curve is the fit including the value for 1988. The solid and dashed
curves in (b) stand for $r(t)$ evaluated with Eq.\ (\ref{r_time}) of
the NLF model, for the shorter and longer series, respectively. In both
drawings, the vertical lines indicate the corresponding values of
$t_c$.}
\end{figure*}

Economic growth was uneven in the 1980's \cite{ocampo90,ocampo91}.
After the end of the civil war the restructure and rebuild of the
economy lead to a positive jump of GDP of about 5 percent in 1980 and
1981. However, each year from 1984 to 1990 showed a drop in the GDP.
Reasons for the contraction included the reluctance of foreign banks to
offer new loans, the diversion of funds to fight the new insurrection
against the government hold by the Contras. Daniel Ortega began his
six-year presidential term on January 1985, and established an
austerity program to lower inflation. After the United States Congress
turned down continued funding of the Contras in April 1985, the Reagan
administration ordered a total embargo on United States trade with
Nicaragua. The United States was formerly Nicaragua's largest trading
partner. The government spend a lot money to finance the war against
the Contras during the second part of the 80's decade. The gap between
decreasing revenues and mushrooming military expenditures was filled by
printing large amounts of paper money. Inflation skyrocketed, peaking
at 13,109$\%$ percent annually at the end of 1987. In the last forty
years, at Latin American level, this situation may be only compared
with that occurred in Bolivia in 1985, where the value 11,150$\%$ was
reached. So it sounds like pretty much the same story as many other
hyperinflation events: War debts and foreign pressures inspire
government to print large amounts of money. In 1988 began the efforts
to stop the spiral of hyperinflation, however, the success of this
attempt was very poor \cite{ocampo90,ocampo91}. It is worthwhile to
notice that in the Nicaragua's administration no Central Bank existed.
In the February 1990 Violeta Barrios de Chamorro won the elections
defeating Ortega. Afterwards, the usual stabilization procedures were
rigorously applied leading to a stable regime in 1992.

Sornette {\it et al.} \cite{sornette03} analyzed the CPI from 1969 to
1991 evaluated with data taken from Ref.\ \cite{imf}, the corresponding
values of $i(t)$ for the regime close to the hyperinflation's peak are
listed in Table \ref{tab:Nicaragua}. Using the same procedure as that
applied in the case of Brazil these authors fitted $P(t)$ instead of
$\ln P(t)$ with Eq.\ (\ref{price1}). The obtained results are included
in Table \ref{tab:table1}.

The inflation occurred in Nicaragua during the 1980's decade is
revisited in the present work. In so doing, we found revised data of
$i(t)$ published by the IMF in the section of Economy in Ref.\
\cite{imf_2011}, which are also included in Table \ref{tab:Nicaragua}.
A glance at this table indicates that the values at the beginning and
at the end of both series are equal. However, the new values forming
the peak of hyperinflation are shifted one year towards prior date
and, in addition, the years corresponding to the values of about
7000$\%$ and 3000$\%$ are interchanged. The latter feature changes the
profile of the crossover to the stationary regime. The CPI and GRI
evaluated with data taken from Ref.\ \cite{imf_2011} are displayed in
panels (a) and (b) of Fig.\ \ref{fig:Nicaragua}, respectively. The
results plotted in Fig.\ \ref{fig:Nicaragua}(b) indicate that the
inflation measured in 1989 can be assigned to the beginning of the
decrease towards the stable regime. Therefore, we analyzed the
hyperinflation using two sets of data, one considering values of CPI
from 1969 to 1987 and the other including the value of 1988 also. This
sort of fits yielded the parameters listed in Table~\ref{tab:table1}.
The solid curve in Fig.\ \ref{fig:Nicaragua}(a) indicates the fit of
the shorter series to Eq.\ (\ref{price1}). A very steep slope of the
CPI may be observed close to $t_c$. If the value of 1988 is included
in the analysis, a piece of information on stabilization is taken into
account, then the fit to Eq.\ (\ref{price1}) gives, as expected, a
larger critical time. This value together with the other parameters
and the $\chi$ are included in Table \ref{tab:table1} and the fit is
depicted by a dashed curve in Fig.\ \ref{fig:Nicaragua}(a). For both
sets of parameters the GRI is evaluated with Eq.\ (\ref{r_time}). The
results are displayed in Fig.\ \ref{fig:Nicaragua}(b), where a good
matching with measured data may be observed. In addition, in both
panels of Fig.\ \ref{fig:Nicaragua} the vertical lines stand for the
determined values of $t_c$.

\begin{figure*}
\includegraphics[width=8cm, height=6cm]{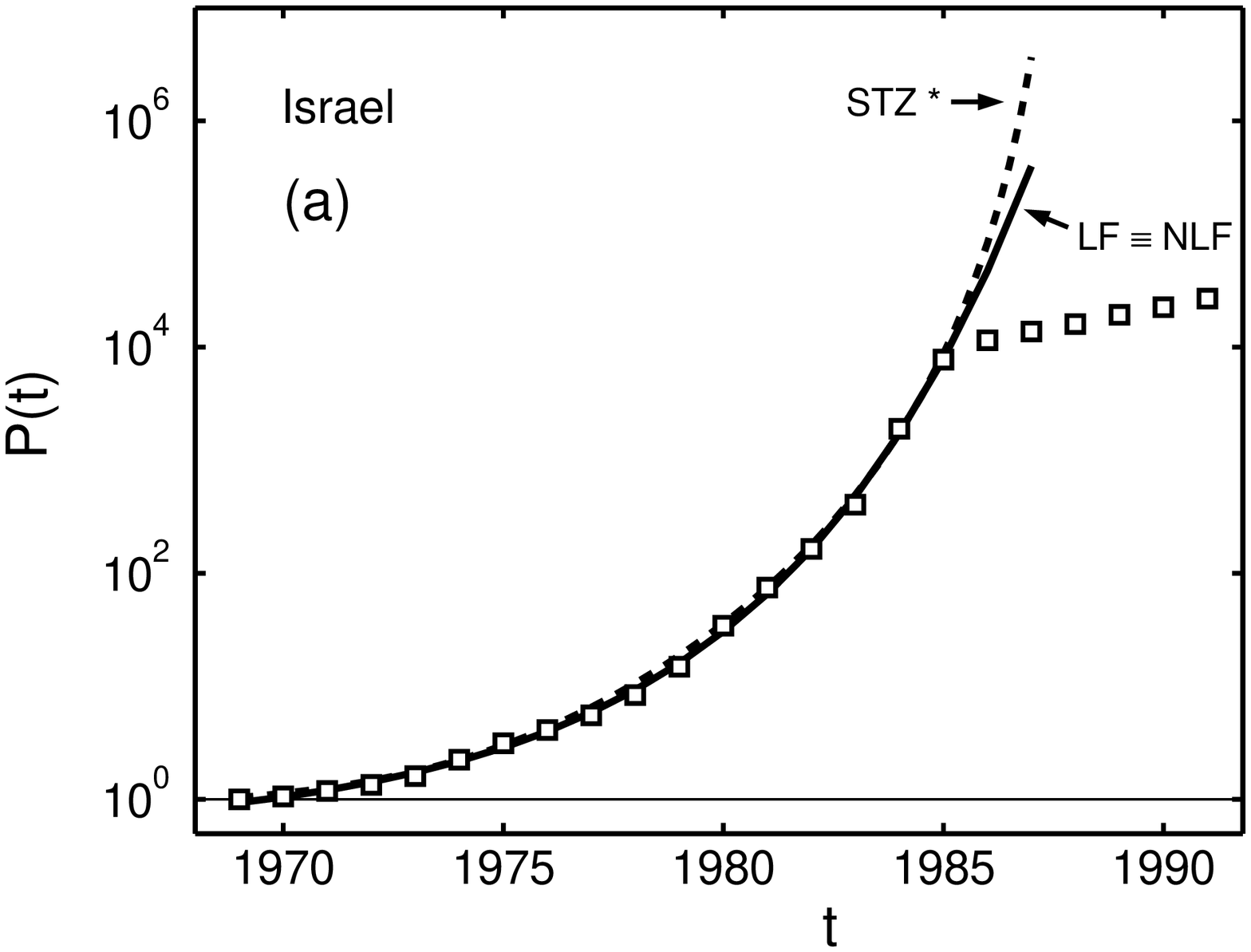}
\includegraphics[width=8cm, height=6cm]{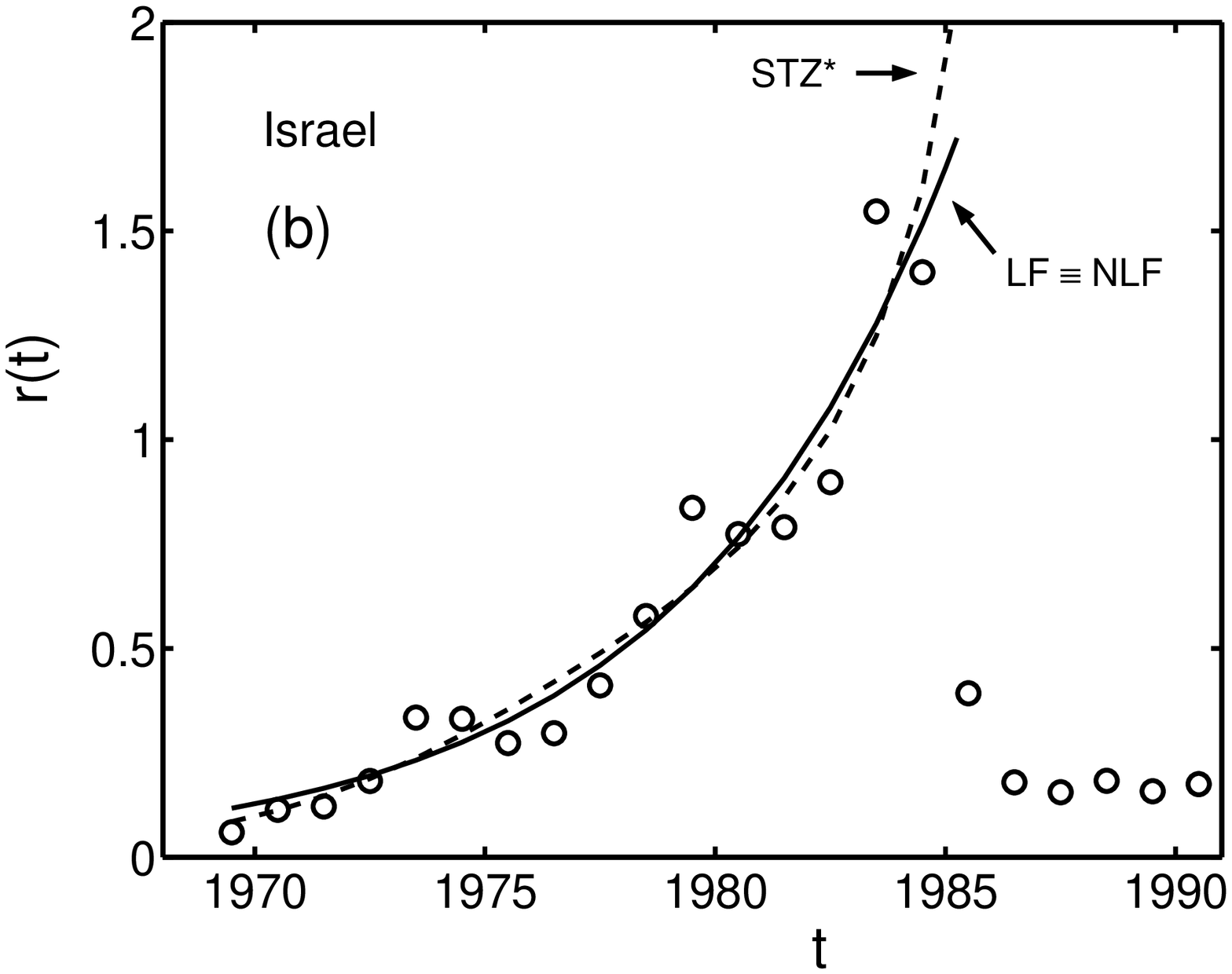}
\caption{\label{fig:Israel_h}(a) Squares are yearly CPI in Israel
since 1969 to 1991, normalized to P($t_0$=1969)=1, presented in a
semi-logarithmic plot. The solid curve indicates the fit of $\ln P(t)$
since 1969 to 1985 to Eqs.\ (\ref{p_mizu}) and (\ref{price1})
corresponding to the LF and NLF models, respectively, while the dashed
curve stands for the fit of $P(t)$ to Eq.\ (\ref{price1}) as reported
by STZ* (see text). (b) Circles are GRI for the same period as in (a).
The solid curve was evaluated with Eqs.\ (\ref{r_mizu}) and
(\ref{r_time}) provided by LF and NLF models, respectively, while the
dashed curve was computed with Eq.\ (\ref{r_dP_1}) of the STZ*
procedure (see text).}
\end{figure*}

\begin{figure}
\includegraphics[width=8cm, height=6cm]{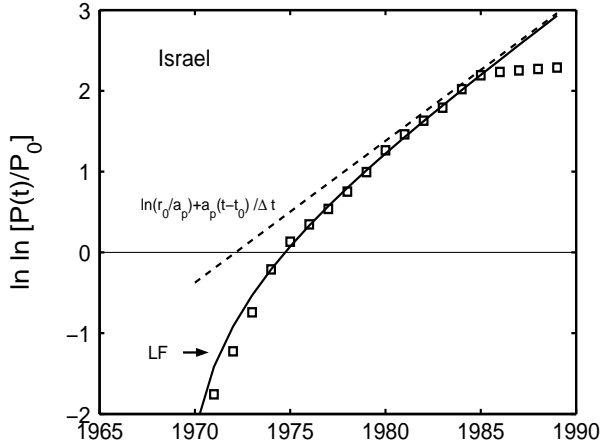}
\caption{\label{fig:Israel_lnCPI} Squares are yearly CPI data in
Israel since 1970 to 1989 taken as $\ln \ln [P(t)/P_0]$. The solid
curve is the fit of data since 1969 to 1985 with the complete Eq.\
(\ref{p_mizu}), while the dashed straight line is the asymptotic
behavior of the LF model, i.e. $\ln \ln [P(t)/P_0]_{asympt}$, given by
Eq.\ (\ref{lnlnP}).}
\end{figure}

\subsection{A drawback of the NLF model: Israel}
\label{sec:drawback}

Let us now focus on the case of Israel. The difficulties for
determining a reasonable $t_c$ from data of this country have a
different origin from those found in the cases of Brazil and Nicaragua.
Figure\ \ref{fig:Israel_h}(a) shows the yearly data for the CPI in
Israel computed using data taken from a Table of the International
Monetary Fund (IMF) \cite{imf}. The evolution of this CPI may be
summarized as follows. Deterioration of the internal and external
conditions following the energy crisis and the Yom Kippur War of 1973
led to an increase in inflation. The labor government chose to
accommodate it in the same way as Brazil. The single-digit rates of
inflation in the 1960's, developed to an annual inflation rate of about
$40\%$ in 1974-75, about $80\%$ in 1978, and got triple-digit rates of
about $400\%$ at their peak in the mid-1980's. Leiderman and Liviatan
\cite{leiderman03} attributed this response to the implicit preference
for short-term considerations of avoiding unemployment over long-term
monetary stability. In 1985 a new strategy was applied that combines
drastic cuts in government deficit and fixed nominal variables
(anchors), i.e. the exchange rate, wages and bank credit. This approach
succeeded in bringing down inflation to a moderate level (near $10\%$).
In the 90's inflation targeting was adopted and inflation came down to
levels recommended by the Organization for Economic Cooperation and
Development (OECD), i.e., about 2 or $3\%$.

The hyperinflation since 1969 to 1985 clearly exhibits a faster
than exponential growth as indicated by the upward curvature of
the logarithm of CPI as a function of time displayed in Fig.\
\ref{fig:Israel_h}(a). In a first step, we fitted data of CPI to
Eq.\ (\ref{price1}) in a similar way to that performed by STZ
\cite{sornette03}. The obtained parameters and the $\chi$ are
listed in Table \ref{tab:table1}. A glance at this table indicates
a critical time $t_c=2061$ and a small exponent of the power law
$\beta=0.069$ ($\alpha \simeq 13$). According to STZ
\cite{sornette03} both these values are unrealistic for a
developing hyperinflation, in addition, they state that the
results are not improved by reducing the time intervals over which
the fits are performed. In addition, they attributed these
problems to the fact that the later prices close to the end of the
time series start to enter a cross-over to a saturation.

Furthermore, the values $t_c = 2061 \pm 72$ and $\beta=0.069\pm
0.061$, which agree with that mentioned by STZ \cite{sornette03},
were obtained by stopping the minimization procedure when the
variation of $\chi^2$ between the $i+1$ and $i$ iterations was
smaller than a standard choice $10^{-1}\,\%$. However, if one
allows to continue the iterations a correlation between these both
parameters becomes clear, $t_c$ increases while $\beta$ decreases
approaching zero, this happens in such a way that the product
$\beta \times (t_c-t_0)$ converges to a constant yielding a well
defined value of the parameter $a_p$ given by Eq.\
(\ref{a_p_lim}). For instance, in Table \ref{tab:table1} we quoted
values obtained when the change of $\chi^2$ becomes less than
$10^{-3}\,\%$.

Let us now show that by following the route $\beta \to 0$
described by numerical minimization the NLF expressions for GRI
and CPI converge to Eqs.\ (\ref{r_mizu}) and (\ref{p_mizu})
derived in the MTT's LF model \cite{mizuno02}, which corresponds
to set $\beta=0$ in Eq.\ (\ref{rate42}). The expression for
$r(t,\beta \to 0)$ is obtained starting from Eq.\ (\ref{r_time})
\begin{eqnarray}
r(t,\beta \to 0) &=& r_0\,\lim_{\beta \to 0}\,\biggr[\frac{1}
{1 - \beta\,a_p\,r_0^\beta \left(\frac{t-t_0}{\Delta t}\right)}
\biggr]^{1/\beta} \nonumber\\
&=& r_0\,\lim_{\beta \to 0}\,\biggr[ \frac{1}{1-\beta\,a_p\,
\left(\frac{t-t_0}{\Delta t}\right)} \biggr]^{1/\beta} \;.
\label{r_time_i}
\end{eqnarray}
It is noteworthy that after the change of variable $\beta=q-1$ the
last expression can be identify with the limit $q \to 1$ of the
{\it q-exponential function}, i.e. $e^x_q$, used in studies of
nonextensive statistical mechanics and economics \cite{tsallis03} 
\begin{eqnarray}
r(t,\beta \to 0) &=& r(t,q \to 1) \nonumber\\
&=& r_0\,\lim_{q \to 1}\,\biggr[ \frac{1}{1-(q-1)\,a_p\,
\left(\frac{t-t_0}{\Delta t}\right)} \biggr]^{1/(q-1)}
\nonumber\\
&=& r_0\,\lim_{q \to 1}\,\biggr[e_q^{[a_p\,(t-t_0)/\Delta t]}\,
\biggr] \nonumber\\
&=& r_0\,\exp \biggr[ a_p\,\left(\frac{t-t_0}{\Delta t}\right)
\biggr] \;, \label{r_time_b}
\end{eqnarray}
because $e_1^x=e^x$ (see Ref.\ \cite{tsallis03}). Furthermore,
imposing the limit $\beta \to 0$ in Eq.\ (\ref{lptn}) for CPI one
gets
\begin{eqnarray}
&&p(t, \beta \to 0) = p_0 \nonumber\\
&&+ \lim_{\beta \to 0}\,\biggr\{ \frac{r_0^{1-\beta}}{(1-\beta)
\,a_p}\biggr( \biggr[\frac{1}{1-\beta\,a_p\,r_0^\beta
\left(\frac{t-t_0}{\Delta t}\right)}
\biggr]^{\frac{1-\beta}{\beta}} - 1 \biggr) \biggr\} \nonumber\\
&&= p_0 + \frac{r_0}{a_p} \biggr\{ \lim_{\beta \to 0}
\biggr[\frac{1}{1-\beta\,a_p\,r_0^\beta
\left(\frac{t-t_0}{\Delta t}\right)}\biggr]^{\frac{1}{\beta}} - 1
\biggr\} \nonumber\\
&&= p_0 + \frac{r_0}{a_p}\, \biggr\{ \exp \biggr[ a_p\,
\left(\frac{t-t_0}{\Delta t}\right) \biggr] - 1 \biggr\} \;.
\label{p_time_b}
\end{eqnarray}
The results obtained in Eqs.\ (\ref{r_time_b}) and
(\ref{p_time_b}) are equal to the corresponding formulas of the LF
model. Therefore, we also fitted the CPI data for the period
1969-1985 directly with LF's Eq.\ (\ref{p_mizu}). The obtained
parameters together with the r.m.s residue $\chi$ are included in
Table \ref{tab:table1}. The good quality of the fit may be
observed in Fig.\ \ref{fig:Israel_h}(a). Notice the excellent
agreement between the values of $r_0$, $a_p$, and $\chi$ yielded
by the LF approach and those obtained from the ``long'' fit with
Eq.\ (\ref{price1}) of the NLF model. Both these fits are
equivalent as indicated in Fig.\ \ref{fig:Israel_h}(a). It is also
worthwhile to mention that the present value for the parameter
utilized in the MTT description, i.e. $B_{MTT}=1+2a_p=1.35$, is in
good agreement with the result $1.4$ quoted in Table 1 of Ref.\
\cite{mizuno02}. For the sake of completeness we plotted in Fig.\
\ref{fig:Israel_h}(b) the measured data of GRI together with the
theoretical values yielded by Eqs.\ (\ref{r_mizu}) and
(\ref{r_time}) provided by LF and NLF models, respectively.

The analysis was completed by fitting data of CPI from 1969 to
1984, i.e. stopping the series before the imposition of the final
stabilization. The results are also included in Table
\ref{tab:table1}, no sizable differences from the fits to the
larger series were observed.
  
The success of the LF's description in the case of Israel is due
to the fact that a double-exponential law is an upper bound for
data of $P(t)$. This feature is depicted in Fig.\
\ref{fig:Israel_lnCPI}, where measured values of $\ln \ln P(t)$
are plotted together with the fit with the complete LF model and
the straight line given by the asymptotic expression of this model
\begin{equation}
\ln \ln \biggr[\frac{P(t)}{P_0} \biggr]_{asympt} = \ln \biggr(
\frac{r_0}{a_p} \biggr) + a_p\,\biggr(\frac{t-t_0}{\Delta t}
\biggr) \;. \label{lnlnP}
\end{equation}
One may realize that experimental data of the hyperinflation (i.e.
until 1985) approach the asymptotic straight line from bellow.

Although the LF model provides a good fit, it does not predict any
$t_c$ indicative for a possible crash of the economy. In order to
estimate a $t_c$, the authors of Ref.\ \cite{sornette03} adopted
the same trick as that used in the cases of Brazil and Nicaragua,
i.e., fitting data of $P(t)$, rather than values of  $\ln P(t)$,
with the r.h.s. of Eq.\ (\ref{price1}). The results reported in
Table 2 of Ref.\ \cite{sornette03} are included in the present
Table \ref{tab:table1} and the fit is shown in Fig.\
\ref{fig:Israel_h}(a). For the sake of completeness, $r(t)$ was
calculated using Eq.\ (\ref{r_dP_1}) corresponding to the STZ*
choice and displayed in Fig.\ \ref{fig:Israel_h}(b)).
However, this procedure for overhauling the lack of a theoretical
tool able to account for any degree of saturation does not
preserve the logical structure of the entire model. As emphasized
above, $P(t)$ is the exponential of the integral of $r(t)$ as
given by Eq.\ (\ref{pt}). In this case $r(t)$ would be given by
Eq.\ (\ref{r_dP_1}). In turn, this expression for $r(t)$ should be
obtained as a solution of a differential equation, e.g. Eq.\
(\ref{rate00}), which must be formulated in a dynamical
description of this kind of economic system. The latter
requirement is not fulfilled in the analysis performed by STZ*,
hence, it remains as a simple fit to a selected expression only.

\section{Summary and Conclusions}
\label{sec:summary}

In the present work we treated regimes of hyperinflation in
economy. The episodes occurred in Brazil, Israel, and Nicaragua
were revisited. These new studies indicated that after some
management of data outlined in Sec.\ \ref{sec:failures} the cases
of Brazil and Nicaragua were  successfully described within the
frame of the NLF model available in the literature
\cite{sornette03,szybisz09}. This formalism outlined in Sec.\
\ref{sec:feedback} is based on a nonlinear feedback
characterized by an exponent $\beta>0$, see Eq.\ (\ref{rate00}).
In this model, a critical time $t_c$ at which the economy would
blow up can be determined from a finite time singularity of the
form $1/(t_c-t)^{(1-\beta)/\beta}$ exhibited by the CPI.

It was found that the hyperinflation occurred in Brazil from 1969
to 1994 can be satisfactorily well described within the NLF frame
if one assumes that, in fact, there are two successive regimes:
one from 1969 to 1990 previous to the Collor's Plan and the other
subsequent to that plan. For the first regime a reasonable $t_c$
was obtained. This feature is in agreement with the statement of
Lucas \cite{lucas76}, that parameters can change once policy
changes.

On the other hand, the episode developed in Nicaragua can be well
described when the corrected data are considered. The corrections
of the inflation series reported in the literature are centered
around the peak of the data. The corrected values yielded a
reasonable $t_c$ within the frame of the NLF model. 

Finally, by applying the NLF model to the weaker hyperinflation
of Israel no $t_c$ is got. Moreover, the data are consistent with
$\beta \to 0$ and, in turn, this limit leads to the linear
feedback proposed in Ref.\ \cite{mizuno02} which does not predict
any $t_c$. In this case there is neither bifurcation of CPI nor
correction of data, instead, there is a slowly increasing
hyperinflation due to a permanent but incomplete effort to stop
inflation.

Since it would be of interest to estimate a $t_c$ within a
self-consistent theory even in the case of a weak hyperinflation,
we shall propose in a forthcoming work an extension of the NLF
model including the effect of a latent incomplete stabilization.
This purpose will be achieved by introducing a new parameter
acting on all past $r(t)$.

Let us finish emphasizing that these lessons should not be lost,
but instead should be kept in mind to avoid the repetition of that
unpleasing experiences. Moreover, one should always remain the
statement of Keynes \cite{keynes30}, namely that: ``even the
weakest government can enforce inflation when it can enforce
nothing else''.

\begin{acknowledgments}
This work was supported in part by the Ministry of Science and
Technology of Argentina through Grants PIP 0546/09 from CONICET
and PICT 2011/01217 from ANPCYT, and Grant UBACYT 01/K156 from
University of Buenos Aires.
\end{acknowledgments}

\widetext

\end{document}